\documentclass[aip,jcp,reprint,superscriptaddress,letter]{revtex4-1}

\usepackage{graphicx}
\usepackage{amssymb,amsmath}
\usepackage{bm}

\begin{document}

\title{Magnetic dipole transitions in the OH $A\,^2\Sigma^+ \leftarrow X\,^2\Pi$ system} %Title of paper

\author{Moritz Kirste}
\affiliation{Fritz-Haber-Institut der Max-Planck-Gesellschaft, Faradayweg 4-6, 14195 Berlin, Germany}
\author{Xingan Wang}
\affiliation{Fritz-Haber-Institut der Max-Planck-Gesellschaft, Faradayweg 4-6, 14195 Berlin, Germany}
\author{Gerard Meijer}
\affiliation{Fritz-Haber-Institut der Max-Planck-Gesellschaft, Faradayweg 4-6, 14195 Berlin, Germany}
\author{Koos B. Gubbels}
\affiliation{Institute of Theoretical Physics, University of Cologne, Z\"{u}lpicher Str. 77, 50937 Cologne, Germany}
\affiliation{Fritz-Haber-Institut der Max-Planck-Gesellschaft, Faradayweg 4-6, 14195 Berlin, Germany}
\affiliation{Radboud University Nijmegen, Institute for Molecules and Materials, Heijendaalseweg 135, 6525 AJ Nijmegen, The Netherlands}
\author{Ad van der Avoird}
\affiliation{Radboud University Nijmegen, Institute for Molecules and Materials, Heijendaalseweg 135, 6525 AJ Nijmegen, The Netherlands}
\author{Gerrit C. Groenenboom}
\affiliation{Radboud University Nijmegen, Institute for Molecules and Materials, Heijendaalseweg 135, 6525 AJ Nijmegen, The Netherlands}
\author{Sebastiaan Y.T. van de Meerakker}
\affiliation{Radboud University Nijmegen, Institute for Molecules and Materials, Heijendaalseweg 135, 6525 AJ Nijmegen, The Netherlands}
\affiliation{Fritz-Haber-Institut der Max-Planck-Gesellschaft, Faradayweg 4-6, 14195 Berlin, Germany}

%\date{\today}
\date{July 19, 2012}

\begin{abstract} We report on the observation of magnetic dipole allowed transitions in the well-characterized $A\,^2\Sigma^+ - X\,^2\Pi$ band system of the OH radical. A Stark decelerator in combination with microwave Rabi spectroscopy is used to control the populations in selected hyperfine levels of both $\Lambda$-doublet components of the $X\,^2\Pi_{3/2},v=0,J=3/2$ ground state. Theoretical calculations presented in this paper predict that the magnetic dipole transitions in the $\nu'=1 \leftarrow \nu=0$ band are weaker than the electric dipole transitions by a factor of $2.58\times 10^3$ only, i.e., much less than commonly believed. Our experimental data confirm this prediction.
\end{abstract}

\pacs{}% insert suggested PACS numbers in braces on next line

\maketitle 
The hydroxyl radical (OH) plays a central role in many areas of chemistry and physics, and is one of the most extensively studied molecular species to date. In 1950, Meinel discovered that emission from vibrationally excited OH radicals in the Earth's atmosphere is responsible for the infrared night-time air glow \cite{Meinel:1950}. Detection of the 18~cm absorption lines in the radio spectrum of Cassiopeia A by Weinreb et al. in 1963 revealed the presence of OH in interstellar space \cite{Weinreb Nature 1963}. Shortly after, the OH radical was identified as the first molecule to form astrophysical (mega)masers \cite{Weaver Nature 1965,Baan Astrophysical Journal 1982}. Since then, a wealth of spectroscopic investigations has been carried out in the microwave, infrared, and ultraviolet part of the spectrum, unraveling the electronic, vibrational, rotational, and hyperfine structure of the OH radical. 

The OH ($^2\Pi$) radical (together with the similar NO ($^2\Pi$) radical) has also been established as the paradigm for molecular collisions studies. Interest in these open-shell radical species stems from their importance in combustion and atmospheric environments, as well as from their complex rotational structure that exhibits spin-orbit and $\Lambda$-doublet splittings. Ingenious methods have been developed to select OH ($^2\Pi$) radicals in a single rotational (sub)level, to orient them in space \cite{ter Meuler Phys. Rev. Lett. 1976,Hain Chem. Phys. Lett. 1996}, and to tune their velocity \cite{Gilijamse Science 2006 OH+Xe}. These methods have allowed collision experiments of transient species at the fully state-resolved level, and have contributed enormously to our present understanding of how intermolecular potentials govern molecular collision dynamics.

Recently, the OH radical has emerged as a benchmark molecule in the rapidly developing field of cold molecules \cite{Meerakker Chemical Reviews}. The OH radical was one of the first molecular species to be slowed down \cite{Bochinski Phys. Rev. Lett. 2003} and to be confined in traps \cite{Meerakker Phys. Rev. Lett. 2005}. In the near future, comparison of high-resolution spectroscopic data on cold OH radicals in the laboratory with interstellar megamaser observations may reveal a possible time variation of fundamental constants \cite{Hudson Phys. Rev. Lett. 2006}.

\begin{figure}
	\centering
	\includegraphics[width=\columnwidth]{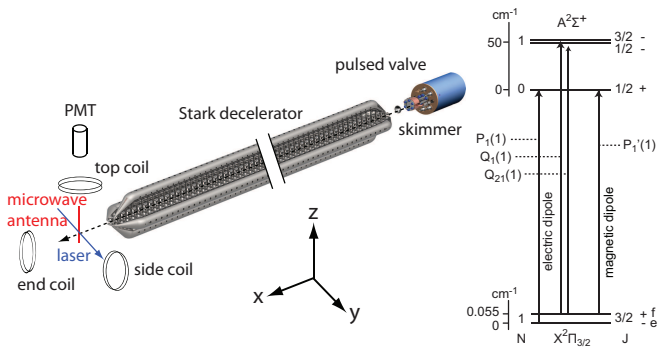}
	\caption{Scheme of the experimental setup. The inset shows the electric dipole allowed and magnetic dipole allowed transition used for the detection of the two $\Lambda$-doublet components in the OH ground state.}
	\label{fig:exp_setup}
\end{figure}

In the vast majority of experiments, ground state OH radicals are detected via laser induced fluorescence (LIF) after optical excitation on electric dipole allowed (EDA) transitions of the $A\,^2\Sigma^+ \leftarrow X\, ^2\Pi$ band using a pulsed dye laser. An important property of the $A-X$ band is that it allows one to selectively probe the population of individual $\Lambda$-doublet components of opposite parity within a rotational state. Although the $\Lambda$-doublet splittings are typically much smaller than the bandwidth of pulsed dye lasers, the measurement of populations in selected $\Lambda$-doublet components is facilitated by the parity selection rules of EDA transitions and the large energy splitting between levels of opposite parity in the $A\, ^2\Sigma^+$ state (see inset to Figure \ref{fig:exp_setup}). Similar schemes are used to probe $\Lambda$-doublet component resolved populations in other $^2\Pi$ molecules such as NO, CH, and SH. 

%Extreme care, however, must be taken when using this approach. In recent experiments in our laboratory, molecular beams of OH with an almost perfect quantum state purity were produced via the Stark deceleration technique, where $\geq 99.999$ \% of OH radicals of the $^2\Pi_{3/2}, J=3/2$ rotational ground state reside in the upper $\Lambda$-doublet component ($f$ parity); the lower $\Lambda$-doublet component of $e$ parity is effectively depopulated in the Stark-deceleration process. When the populations in the $e$ and $f$ components were probed using LIF via the $A\leftarrow X$ transition, however, the apparent population in the $e$ state appeared at least one order of magnitude too large. A spectroscopic re-investigation using a laser with a much narrower bandwidth revealed that magnetic dipole allowed (MDA) transitions were responsible for this effect \cite{Kirste Science 2012}.

Extreme care, however, must be taken when using this approach. In recent experiments in our laboratory, molecular beams of OH with an almost perfect quantum state purity were produced via the Stark deceleration technique. In these experiments, $\geq 99.999$ \% of OH radicals in the $^2\Pi_{3/2}, J=3/2$ rotational ground state reside in the upper $\Lambda$-doublet component of $f$ %parity
symmetry; the lower $\Lambda$-doublet component of $e$ %parity
symmetry is effectively depopulated in the Stark-deceleration process.  When the populations in the $e$ and $f$ components were probed using LIF via the $A\leftarrow X$ transition, however, the apparent population in the $e$ state appeared at least one order of magnitude too large. A spectroscopic re-investigation using a laser with a much narrower bandwidth revealed that magnetic dipole allowed (MDA) transitions were responsible for this effect \cite{Kirste Science 2012}.

Magnetic dipole allowed transitions have rarely been observed in laser excitation spectra of heteronuclear molecules \cite{Yang Journal of Molecular Spectroscopy 2010}. Their existence is generally neglected in quantitative measurements of state populations, potentially leading to a significant misinterpretation of detector signals. In homonuclear molecules, MDA transitions between electronic states are well known to result in ``forbidden'' band systems that violate the rigorous selection rules for electric dipole transitions. The most famous example is the atmospheric oxygen band, which appears in the red part of the solar spectrum. In contrast, MDA transitions in heteronuclear molecules mostly exist as weak satellite lines parallel to strong EDA transitions. The general rule of thumb is that MDA transitions are about a factor $10^5$ weaker than the corresponding EDA transitions \cite{Herzberg}. Already in the 1920's, weak satellite lines in the $A-X$ emission band of OH were observed that appeared to correspond to transitions to the ``wrong'' $\Lambda$-doublet component \cite{Dieke:Nature115:194,Watson:Nature117:157,Mulliken:PhysRev32:388,Jack:PRSA120:222}. These lines were tentatively attributed to the MDA transitions by Van Vleck in 1934 \cite{Vleck:AJ80:161}, but received little attention ever since. 

Here, we present a detailed analysis of MDA transitions in the $A\,^2\Sigma^+ \leftarrow X\,^2\Pi$ band of OH. We show that the satellite MDA transitions are surprisingly strong, and only three orders of magnitude weaker than the main EDA transitions. In our experiment we use a Stark-decelerator to produce packets of OH radicals that reside exclusively in the upper $\Lambda$-doublet component of $f$ %parity
symmetry. A controlled fraction of the population is transferred to the lower component of $e$ %parity
symmetry by using a microwave field. The MDA and EDA $A\,^2\Sigma^+, v=1 \leftarrow X\,^2\Pi, v=0$ transitions originating from the $f$ and $e$ level, respectively, are spectroscopically resolved using a narrowband pulsed dye laser. The observed ratio of the signal intensities agrees well with theoretical calculations for the EDA and MDA transition strengths. 

The relevant energy levels and electronic transitions are shown in the inset to Figure \ref{fig:exp_setup}. The electronic ground state of OH has a $X\,^2\Pi$ configuration. Each rotational level, labeled by $J$, splits into two $\Lambda$-doublet components which are separated by 0.055~cm$^{-1}$ for the $J=3/2$ rotational ground state. The upper and lower components have $+$ and $-$ parity, and are indicated by the spectroscopic labels $f$ and $e$, respectively. Each of the $\Lambda$-doublet components of the $J=3/2$ state is split into $F=1$ and $F=2$ hyperfine levels. The four resulting levels are referred to hereafter as $\left|X,f,+,F=2\right\rangle$, $\left|X,f,+,F=1\right\rangle$, $\left|X,e,-,F=2\right\rangle$ and $\left|X,e,-,F=1\right\rangle$. 

The first electronically excited state of OH has a $A\,^2\Sigma^+$ configuration. In our experiments, only the $N=0, J=1/2$ rotational ground state of $+$ parity is of relevance. This state is split into two hyperfine states $F=0$ and $F=1$ that are separated by 0.026~cm$^{-1}$, and are referred to hereafter as $\left|A,+,F=0\right\rangle$ and $\left|A,+,F=1\right\rangle$. The EDA ($P_1(1)$) and MDA ($P'_1(1)$) $A-X$ transitions couple the $\left|X,e,-\right\rangle$ and $\left|X,f,+\right\rangle$ states to the $\left|A,+\right\rangle$ states following the parity changing and parity conserving selection rules for EDA and MDA transitions, respectively.

Our experimental setup is schematically shown in Figure \ref{fig:exp_setup}. A packet of OH ($X\,^2\Pi_{3/2}, v=0, J=3/2, f$) radicals with a velocity of 448~m/s is produced by passing a molecular beam of OH through a 2.6 meter long Stark decelerator \cite{Scharfenberg PRA 2009 newdec}. The Stark decelerator efficiently deflects molecules in the $\left|X,e,-\right\rangle$ states. A phase angle $\phi_0=50^\circ$ is used to ensure that the OH radicals that exit the decelerator reside exclusively in the $\left|X,f,+,F=2\right\rangle$ state. The end of the Stark decelerator is electrically shielded to prevent any electric stray fields to penetrate into the interaction area.

A controlled fraction of the OH radicals is transferred into the $\left|X,e,-,F=1\right\rangle$ state by inducing the $\left|X,f,+,F=2\right\rangle \rightarrow \left|X,e,-,F=1\right\rangle$ transition at 1.72~GHz with a microwave pulse. For this purpose a 90~mm long microwave antenna is installed 38~mm downstream from the decelerator and perpendicular to the molecular beam axis. No frequency-matched microwave resonator was used. The microwaves are reflected by the vacuum chamber walls filling the whole vacuum chamber, and we assume the microwaves to be unpolarized. The microwave duration and power can be controlled via a microwave switch and attenuator, respectively. The magnetic field in the interaction region is controlled by three copper coils with a diameter of 31~cm each, that are mounted 30~cm from the interaction area. One coil is positioned above the interaction area, one at the side and one at the end.

Two lasers are used to detect the OH radicals via LIF using the 1-0 band of the OH $A\,^2\Sigma^+ \leftarrow X\,^2\Pi_{3/2}$ transition around 282~nm. The first laser, a pulsed dye laser (PDL) with a bandwidth of 1.8~GHz, is used to probe the population in the $\left|X,e,-\right\rangle$ state via the EDA $P_1(1)$ transition. The second laser, a pulsed dye amplifier (PDA) seeded by a single mode ring dye laser, has a bandwidth of ~120 MHz and is used to separate the $P_1(1)$ and $P'_1(1)$ transitions. The power of the PDL and PDA lasers are adjusted to ensure that the transitions are induced under saturated and unsaturated conditions, respectively, and both lasers are linearly polarized in the $z$ direction (see Figure \ref{fig:exp_setup} for the definition of the coordinate system). The off-resonant fluorescence is imaged into a photomultiplier tube (PMT).

\begin{figure}
	\centering
	\includegraphics[width=\columnwidth]{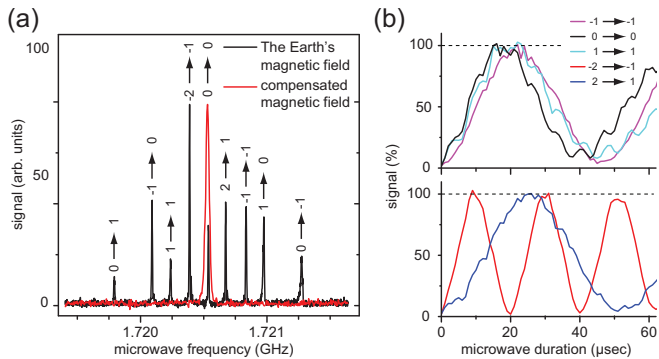}
	\caption{(a): Frequency scan over the $\left|X,f,+,F=2\right\rangle \rightarrow \left|X,e,-,F=1\right\rangle$ transition in the presence of the Earth's magnetic field (black) and the compensated magnetic field (red). Spectra are scaled to each other to fasciliate a better comparison. The nine $\left|X,f,+,F=2,M_F\right\rangle \rightarrow \left|X,e,-,F=1,M_F\right\rangle$ transitions are indicated. (b): Rabi oscillations of selected $\left|X,f,+,F=2,M_F\right\rangle \rightarrow \left|X,e,-,F=1,M_F\right\rangle$ transitions.}
	\label{fig:spectra_rabi_cycles}
\end{figure}

In the presence of a magnetic field, the $F=1$ and $F=2$ hyperfine states split into 3~and 5~$M_{F}$ Zeeman sublevels, respectively, that are readily resolved in the microwave spectrum. This is illustrated in Figure \ref{fig:spectra_rabi_cycles}(a) that shows the $\left|X,f,+,F=2\right\rangle \rightarrow \left|X,e,-,F=1\right\rangle$ spectrum around 1.72~GHz, recorded with the broadband PDL system. In the black spectrum no currents are applied to the coils, and nine transitions can be identified corresponding to the nine allowed $\left|X,f,+,F=2, M_F\right\rangle \rightarrow \left|X,e,-,F'=1, M'_F\right\rangle$ transitions that are split by the Earth's magnetic field. 

For an unambiguous interpretation of the EDA and MDA $A-X$ transitions, and to measure their relative strengths, it is convenient to choose the laser polarization direction parallel to the space quantization axis. The Earth's magnetic field, however, is not suitable for this, as the direction of the magnetic field vector is in general not parallel to the laser polarization axis. We therefore follow the approach to first compensate the Earth's magnetic field by applying currents to the three coils, and then to apply a controlled magnetic field that is parallel to the $z$ axis, i.e., the laser polarization axis. The red curve in Fig.~\ref{fig:spectra_rabi_cycles}(a), shows the microwave spectrum that is recorded when currents of 2.10~A, 1.60~A and 0.35~A are passed through the top, side and end coils, respectively. It is seen that in this configuration the Earth's magnetic field is compensated and the nine lines merge into one. An additional magnetic field in the $z$ direction can be added by changing the current in the top coil, while keeping the current in the other coils constant. We have chosen to reverse the current in the top coil to generate a magnetic field with a magnitude that is twice as large as the $z$-component of the Earth's magnetic field.

A controlled fraction of the population in each of the $\left|X,f,+,F=2, M_F\right\rangle$ levels can be transferred to an individual $M_F$ component of the $\left|X,e,-,F=1\right\rangle$ level by applying a microwave pulse with a controlled pulse duration and power. In Figure \ref{fig:spectra_rabi_cycles}(b), the fluorescence intensity is shown that is measured for five different microwave transitions as a function of the microwave pulse duration. Clear Rabi oscillations are observed, with different Rabi frequencies for each transition due to the differences in transition strength and the unpolarized microwave radiation. These Rabi oscillations were measured for all nine transitions shown in Figure \ref{fig:spectra_rabi_cycles}(a), and for each transition it was observed that the maxima of the oscillations yield equal signal intensity. We thus conclude that the OH radicals that exit the Stark decelerator are equally distributed over the five $M_F$ levels of the $\left|X,f,+,F=2\right\rangle$ state before the microwave field is applied.

\begin{figure}
	\centering
	\includegraphics[width=\columnwidth]{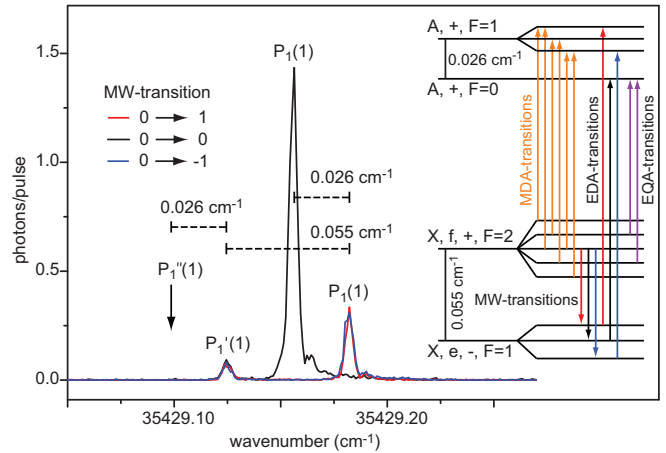}
	\caption{The EDA and MDA OH(A-X) transitions for three different microwave transitions (marked red,black and blue). The corresponding microwave transitions to prepare a population in selected $M_F$ components of the $\left|X,e,-,F=1\right\rangle$ state are shown in the inset and the EDA $P_1(1)$, MDA $P'_1(1)$ and EQA $P''_1(1)$ transitions are indicated. The $\left|X,f,+,F=1\right\rangle$ and $\left|X,e,-,F=2\right\rangle$ levels are not shown in the inset, because the experiment does not populate these levels. The arrow indicates the position of possible EQA transitions $P''_1(1)$.} 
	\label{fig:spectra_result}
\end{figure}

Three different microwave transitions are induced that transfer population from the $\left|X,f,+,F=2,M_F=0\right\rangle$ into the $M_F=1$, $M_F=0$ and $M_F=-1$ levels of the $\left|X,e,-,F=1\right\rangle$ state, respectively. These transitions are indicated by the red, black and blue arrows in the inset in Figure \ref{fig:spectra_result}. For each transition, the microwave pulse duration and power was carefully chosen to transfer (2.5$\pm$1)\% of all molecules from the $\left|X,f,+,F=2,M_F=0\right\rangle$ level. Since this $M_F=0$ level contains one fifth of all $F=2$ molecules, 99.5$\pm$0.2\% of the OH radicals remain in the $\left|X,f,+,F=2\right\rangle$ state, in all three cases. The error (2$\sigma$) is given by the statistical spread of the Rabi oscillations.

The EDA $P_1(1)$ and MDA $P'_1(1)$ $A-X$ transitions are then investigated in these three cases by probing the populations in the $\left|X,e,-\right\rangle$ and $\left|X,f,+\right\rangle$ states with the narrowband PDA system. This laser can spectroscopically resolve the $\Lambda$-doublet splitting in the $\left|X \right\rangle$ state and the hyperfine splitting in the $\left|A\right\rangle$ state, but not the hyperfine splittings in both $\left|X \right\rangle$ states or any Zeeman splittings. For parallel laser polarization and magnetic field direction, both the EDA and the MDA transitions obey the hyperfine selection rule $\Delta F$=0,$\pm$1. The EDA transition has the additional selection rule $\Delta M_F$=0 (with $\Delta F \neq 0$ for $M_F=0$), while MDA transitions can only couple states with $\Delta M_F$=$\pm$1. As indicated in Figure \ref{fig:spectra_result}, there are thus six MDA transitions and only one EDA transition for each case.

In Figure \ref{fig:spectra_result} the MDA $P'_1(1)$ and the EDA $P_1(1)$ transitions are shown that are recorded in the three cases. The MDA $P'_1(1)$ transitions appear at the same position and with equal intensity in all spectra. Depending on the $M_F$ level that is populated in the $\left|X,e,-,F=1\right\rangle$ state, the EDA $P_1(1)$ transition either couples to the $\left|A,+,F=0\right\rangle$ (for $M_F=0$) or the $\left|A,+,F=1\right\rangle$ state (for $M_F=\pm 1$). These transitions are clearly resolved in the spectra. The former transition appears four times more intense than the latter two transitions that are of equal intensity, as is expected theoretically \cite{supplement}. The $\Lambda$-doublet splitting is also recognized. 

Having observed the MDA transition one might wonder about the presence of electric quadrupole allowed (EQA) transitions. For parallel laser polarization and magnetic field direction, an EQA transition can couple states with $\Delta F$=$\pm$2, $\Delta M_F = \pm 1$. In the experiment no EQA $\left|A,+,F=0\right\rangle \leftarrow\left|X,f,+,F=2\right\rangle$ transition was observed, indicating that EQA transitions in the OH (A-X) band are at least two orders of magnitude weaker than MDA transitions. This finding is supported by the theoretical estimate of the EQA transition strength \cite{supplement}. 

The relative strength of the $A-X$ MDA and the EDA transitions can be deduced from the measured spectra, and compared to theory. The strengths of the transitions are calculated from the magnitude of the two transition dipole moments, given by $\mu_{\rm el/mag}=\left| \left\langle A\,^2\Sigma^+,v=1 \right| \bm{\hat{\mu}}_{\rm el/mag} \left| X\,^2\Pi,v=0\right\rangle \right|$ \cite{supplement}. We find $\mu_{\rm el}=0.0525$~a.u. and $\mu_{\rm mag}=0.142$~a.u. for the electric and magnetic transition dipole moments, respectively, so that $\frac{1}{\alpha^2} \cdot \mu_{\rm el}^2/\mu_{\rm mag}^2=2.58\cdot10^3$. Here, $\alpha$ is the fine-structure constant accounting for the relative strength of the magnetic field compared to the electric field of the laser. Magnetic dipole transitions in the OH ($A-X$) band are thus only three orders of magnitude weaker than electric dipole transitions. 

Taking into account the experimental initial distribution of molecules over the quantum states, as well as the direction of the laser polarization and the magnetic field, we find a theoretical ratio of 25.8 for the fluorescence intensities of the EDA $\left|A,+,F=0\right\rangle \leftarrow \left|X,e,-,F=1,M_F=0\right\rangle$ transition and the combined six MDA $\left|A,+,F=1\right\rangle \leftarrow \left|X,f,+,F=2\right\rangle$ transitions \cite{supplement}. The uncertainty in this ratio is estimated to be about 10\% \cite{supplement}. This value agrees well with the experimental value of (18$\pm$8), obtained by comparing the strong central with the left peak in Figure \ref{fig:spectra_result}. The experimental error is mainly given by the statistical error of the population transfer in the microwave field from the $\left|X,f,+,F=2,M_F=0\right\rangle$ to the $\left|X,e,-,F=1,M_F\right\rangle$ levels. 

In this work we reported on the direct measurement of magnetic dipole transitions in laser excitation spectra of the OH $A\,^2\Sigma^+,v=1 \leftarrow X\,^2\Pi_{3/2}, v=0$ band. These satellite transitions appear only three orders of magnitude weaker than the corresponding main electric dipole transitions, and can potentially lead to a misinterpretation of detector signals when the $\Lambda$-doublet-resolved state populations in OH ($X\,^2\Pi$) are measured. This finding may seem of limited significance in some experiments; in experiments in which large differences in $\Lambda$-doublet populations are expected it may be essential. In particular in state-of-the-art molecular beam experiments with unprecedented state purity and precision, magnetic dipole transitions should be carefully considered.

The authors are grateful to Samuel Meek, Nicolas Vanhaecke, Janneke Blokland, Boris Sartakov and Christian Schewe for the support in the realization of the experiment and the analysis of the data. K.B.G. and G.M. acknowledge support from the ERC-2009-AdG under grant agreement 247142-MolChip. K.B.G. and A.v.d.A acknowledge support from the Alexander von Humboldt Foundation. S.Y.T.v.d.M. acknowledges support from the Netherlands Organisation for Scientific Research (NWO) via a VIDI grant. Supplementary information accompanies this paper.

\end{document}

\title{Dipole moment transitions in OH: theory}
%\date{\today}
\date{July 19, 2012}

\begin{abstract}

We study theoretically electric and magnetic dipole transitions in the
OH radical from the electronic, vibrational, and rotational ground state
($X\,^2\Pi_{3/2}, J=3/2, v= 0$) to the first electronically and
vibrationally excited state ($A\,^2\Sigma^{+}, J=1/2, v= 1$).

\end{abstract}

\maketitle

\section{Experimental situation}\label{sec:exp}

We consider Stark-decelerated molecules in the $X\,^2\Pi_{3/2}, J=3/2,
v= 0, f$ state, where $X\,^2\Pi_{3/2}$ designates the electronic ground
state of OH for which $|\Omega|=3/2$ is a nearly good quantum number.
This is also called the $F_1$ spin-orbit manifold in the main text.
Moreover, $J$ is the eigenvalue of the angular momentum operator $
\hat{{\bm J}} = \hat{{\bm L}} +\hat{{\bm S}}+\hat{{\bm R}}$ with
$\hat{{\bm L}}$ the electronic orbital angular momentum, $\hat{{\bm S}}$
the electronic spin, and $\hat{{\bm R}}$ the nuclear rotation, while $v$
is the vibration quantum number and $\Omega$ is the projection quantum
number of $\hat{{\bm J}}$ on the OH bond axis. Using microwave
transitions also the $X\,^2\Pi_{3/2}, J=3/2, v= 0, e$ state is
populated, where the $e$ and $f$ labels denote the so-called
spectroscopic parity. We are interested in magnetic and electric dipole
transitions from the two rovibronic ground states of different parity to
the first electronically excited $A\,^2\Sigma^{+}, J'=1/2, v'= 1$ state.
Since the latter state has $+$ parity upon inversion,
electric dipole transitions are possible from the $e$ ground state with
$-$ parity and magnetic dipole transitions are possible from the
$f$ ground state with $+$ parity. Since the nuclear spin of OH is given
by $I=1/2$, we have a total angular momentum $F=1$ or $F=2$ prior to the
dipole transitions, and $F=0$ or $F=1$ after the transitions.

As explained in the main text, great care was taken experimentally to control
the polarization of the laser and the direction of the externally
applied magnetic field. In the laboratory frame the $x$
direction is parallel to the laboratory floor in the direction of the
molecular beam. The $z$ direction is the vertical direction in the
laboratory, and in this direction the PMT is mounted. The
right-handedness of the coordinate system then fixes the $y$ axis. The
magnetic field and the laser polarization are along the $z$-axis, while
the propagation of the laser is in the $y$ direction, see also Fig. 1 of
the main text. After the Stark deceleration process, the molecules are
initially all in the $X\,^2\Pi_{3/2}, J=3/2, v= 0, f, F=2$ state with
the five $M_F$ states equally populated. These states are thus populated
with a fraction of 20\%. Next, 2.5\% of the molecules in the $M_F=0$
state (i.e., 0.5\% of the total amount of decelerated molecules) is
pumped via a microwave transition to one of the three $X\,^2\Pi_{3/2},
J=3/2, v= 0, e, F=1, M_F$ states. All three possibilities, namely
$M_F=0,\pm 1$, are separately studied experimentally.

After the pumping, the electric and the magnetic dipole transitions are
investigated. The magnetic dipole moment transitions take place from the
$X\,^2\Pi_{3/2}, J=3/2, v= 0, f, F=2, M_F$ states, of which four levels
$(M_F=\pm 1, \pm 2)$ are occupied by 20\% of the molecules and one level
$(M_F=0)$ is occupied by 19.5\%. The electric dipole moment transitions
take place from a $X\,^2\Pi_{3/2}, J=3/2, v= 0, e, F=1, M_F$ state, for
which we distinguish three different experimental cases. Case 1: 0.5\%
of the molecules are in the $M_F=0$ state, case 2: 0.5\% of the
molecules are in the $M_F=1$ state, case 3: 0.5\% of the molecules are
in the $M_F=-1$ state. In all three cases, the other $M_F$ states of the
ground state with spectroscopic $e$ parity are unpopulated. Next, we show that, as a result, there are two independently measured intensity
ratios. Namely, the ratio of the two different electric dipole
transitions (case 2 and 3 have the same strength), and the ratio between
the magnetic and the electric dipole transition strength. In the next
sections, we calculate these ratios from first principles.

\section{Theory}\label{sec:theory}

The transition dipole moments were calculated with the MOLPRO
\cite{molpro:10} program package at the internally contracted
multireference singles and doubles excitation configuration interaction
(icMRSDCI) level. The orbitals were obtained with a state-averaged
complete active space self-consistent-field (CASSCF) calculation
employing the aug-cc-pVQZ one electron basis \cite{dunning:89}. The
active space contained the full valence space, the (1s) core orbital on
the O atom and extra MOs of each symmetry. The C$_{2v}$ point group symmetry
was used and the active space consisted of six $a_1$, three $b_1$, three
$b_2$, and three $a_2$ orbitals. To determine the magnetic transition
dipole moment, we calculated the matrix elements of the magnetic dipole
moment operator $\hat{\boldsymbol{\mu}}_{\rm mag}=-\mu_{\rm B}\hat{{\bm
L}}$ with the Bohr magneton $\mu_{\rm B}=1/2$ in atomic units. The spin
operator $\hat{{\bm S}}$ does not couple a $\Sigma$ state to a $\Pi$
state and hence does not contribute to the magnetic dipole transition moment.
Since the transition dipole moments depend on the OH bond length
$r$, we used a grid of 37 points ranging from $r=1.2~a_0$ to
$r=2.8~a_0$, with extra points lying around the center of the grid. In
order to determine the matrix elements for the transition dipole
moments, we also needed the vibrational ground state of the $X^2\Pi$
potential, and the first vibrationally excited state of the
$A^2\Sigma^+$ potential. We used the electronic potentials determined by
Van der Loo and Groenenboom \cite{loo:07}, and calculated the
vibrational states using the discrete variable representation (DVR)
method based on sinc-functions \cite{groenenboom:93}. Having performed
all these calculations, we were able to determine the electric
transition dipole moment in the body-fixed frame $\langle \mu^{\rm
bf}_{\rm el}\rangle$ (the frame with its $z$ axis along the OH bond
vector), namely
\begin{equation}
\langle \mu^{\rm bf}_{\rm el}\rangle\equiv|\langle A\,^2\Sigma^{+}, v'=
1| \hat{\boldsymbol{\mu}}_{\rm el} |X\,^2\Pi, v= 0\rangle |.
\end{equation}
The magnetic transition dipole moment $\langle \mu^{\rm bf}_{\rm
mag}\rangle$ is defined in a similar way. We find that $|\langle
\mu^{\rm bf}_{\rm el}\rangle| = 0.05249$ and $|\langle \mu^{\rm bf}_{\rm
mag}\rangle| = 0.1417$ in atomic units. The body-fixed transition dipole
moments only have components perpendicular to the OH bond axis for the
$\Pi \rightarrow \Sigma$ transition. Introducing the projection of the
electric transition dipole moment on the body-fixed $z$ axis,
${\mu}^{\rm bf}_{{\rm el},k}$, we have that
\begin{equation}\label{eq:mu}
{\mu}^{\rm bf}_{{\rm el},\pm1}=i \frac{ \langle \mu_{\rm
el}\rangle}{\sqrt{2}} \qquad {\rm and} \qquad {\mu}^{\rm bf}_{{\rm
mag},\pm1}=\mp \frac{ \langle \mu_{\rm mag}\rangle}{\sqrt{2}},
\end{equation}
and in both cases ${\mu}^{\rm bf}_{{\rm el/mag},0}=0$.

In the experiment, care is taken to polarize the laser in the
space-fixed laboratory frame. As a result, in order to treat the
interaction of the transition dipole moments with the laser field, we
need to transform the transition dipole moments from the body-fixed to
the space-fixed frame. The corresponding space-fixed transition dipole
moments are given by
\begin{equation}
{\mu}^{\rm sf}_{{\rm el/mag},m}=\sum_{k}{\mu}^{\rm bf}_{{\rm el/mag},k} D^{(1)*}_{mk}(\phi,\theta,0),
\end{equation}
with $\phi$ and $\theta$ the polar angles of the OH bond axis in the
space-fixed frame, and $D^{(j)}_{mk}(\alpha,\beta,\gamma)=e^{-i m
\alpha}d^{(j)}_{mk}(\beta)e^{-ik\gamma}$ with $d^{(j)}_{mk}(\beta)$ the
Wigner $d$ functions. To proceed, we first consider basis states for the
OH radical that are not yet parity adapted and not mixed due to the
presence of spin-orbit coupling. We consider the states
$|\Lambda,S,\Omega,J,I,F,M_F\rangle$ with $\Lambda$ the projection of
the electronic angular momentum on the body-fixed axis, where $J$ and $I$
are coupled to $F$ according to
\begin{eqnarray}
|\Omega,J,I,F,M_F\rangle = \sum_{M_I,M_J} |J,\Omega,M_J\rangle |I,M_I\rangle \langle J,M_J,I,M_I |F,M_F\rangle,
\end{eqnarray}
where $M_J$, $M_I$, and $M_F$ are the projections of the corresponding
angular momenta on the space-fixed $z$ axis. Moreover, $\langle
J,M_J,I,M_I |F,M_F\rangle$ are the Clebsch-Gordan coefficients. The
rotational wave functions are given by
\begin{eqnarray}
&& |J,\Omega,M_J\rangle =\sqrt{\frac{2J+1}{4\pi}}D^{(J)*}_{M_J,\Omega}(\phi,\theta,0).
\end{eqnarray}
For these states, we find
\begin{eqnarray}
\langle J',\Omega',M_J' |D^{(1)*}_{m k}|J,\Omega,M_J \rangle
=\sqrt{\frac{2 J +1}{2 J' +1}} C^{J',\Omega'}_{J,\Omega,1,k}
C^{J',M'_{J}}_{J,M_{J},1,m},
\end{eqnarray}
with the shorthand notation $C^{F,M_{F}}_{J,M_{J},I,M_I} = \langle J,M_J,I,M_I |F,M_F\rangle$. Using the following relation
\begin{eqnarray}
\sum_{M_J,M'_{J},M_I}
C^{J',M'_{J}}_{J,M_{J},1,m} C^{F,M_F}_{J,M_J,I,M_I}
C^{F',M'_{F}}_{J',M'_{J},I,M_I} &=&\sqrt{(2 J' +1)(2 F+1)}
(-1)^{I+F+J'+1} C^{F',M_F'}_{F,M_F,1,m}
      \begin{Bmatrix}
         J & I & F \\
         F' & 1 & J' \\
      \end{Bmatrix},
\end{eqnarray}
with the curly brackets denoting the 6-$j$ symbol, we obtain
\begin{eqnarray}
\label{eq:tdm}
\lefteqn{\langle ^2\Sigma^+,J',I',F',M'_F| \mu^{\rm sf}_{{\rm el/mag},m}
|^2\Pi_{\Omega},J,I,F,M_F\rangle = \delta_{I',I}
(-1)^{I-J+J'+1-\Omega'-M_F'}
} \nonumber \\
&&
\mbox{}\times\sqrt{(2 J +1)(2 J' +1)(2 F +1) (2 F' + 1)}
\begin{Bmatrix}
         J & I & F \\
         F' & 1 & J' \\
  \end{Bmatrix}
\sum_{k}  {\mu}^{\rm bf}_{{\rm el/mag},k}
  \begin{pmatrix}
        J & 1 & J' \\
        \Omega & k & -\Omega' \\
  \end{pmatrix}
  \begin{pmatrix}
        F & 1 & F' \\
        M_F & m & -M'_F \\
  \end{pmatrix},
\end{eqnarray}
with the round brackets denoting the 3-$j$ symbol. Equation
(\ref{eq:tdm}) is valid for both the electric and the magnetic
transition dipole moment.

For the parity adaptation, we consider first the $\Pi$ state with $J=3/2$ and $I=1/2$, namely
\begin{eqnarray}
|^2\Pi_{\pm \Omega},F,M_F\rangle=|\Lambda= \pm 1, S=1/2 ,\pm \Omega,J=3/2,I=1/2;F,M_F\rangle,
\end{eqnarray}
where $\Omega= 1/2$ or 3/2. The parity adaptation is then performed as follows,
\begin{eqnarray}\label{eq:pief}
|^2\Pi_{|\Omega|},F,M_F, \epsilon=\pm\rangle=(|^2\Pi_{\Omega},F,M_F\rangle \pm |^2\Pi_{-\Omega},F,M_F\rangle)/\sqrt{2},
\end{eqnarray}
where the $\epsilon=+$ corresponds to spectroscopic $e$ parity, while the $-$ corresponds to spectroscopic $f$ parity. 
The relative sign in the wave function, $\epsilon$, is related to the parity under inversion
$p$ by $p=\epsilon (-1)^{J-S}$. Due to the large and negative spin-orbit
constant $A=-139.73$ cm$^{-1}$, $\Omega$ is almost a good
quantum number, and the state with approximately $\Omega=3/2$ is
the ground state. We denote the latter by $F_1 {^2\Pi}_{3/2}$, and it is given
by
\begin{eqnarray}\label{eq:pi}
|F_1 {^2\Pi}_{3/2},F,M_F, \epsilon \rangle=c_{1/2} |^2\Pi_{1/2},F,M_F,\epsilon\rangle + c_{3/2} |^2\Pi_{3/2},F,M_F,\epsilon\rangle,
\end{eqnarray}
where $c_{1/2} = \sqrt{(X+Y-2)/2X}$ and $c_{3/2} = \sqrt{(X-Y+2)/2X}$
with $X = \sqrt{4(J+1/2)^2+Y(Y-4)}$ and $Y = A/B$, see e.g.\ \cite{bernath:05}.
For the rotational constant, we used $B=18.515$
cm$^{-1}$. The final state in the experiment is of $+$ parity. Moreover,
for this state we have $J'=1/2$ and $I'=1/2$, so that $F'=0$ or 1. The
corresponding parity adapted wave function is given by
\begin{eqnarray}\label{eq:sigma}
| ^2\Sigma^+,F',M'_F,+\rangle = (|^2\Sigma^+_{1/2},F',M'_F\rangle+|^2\Sigma^+_{-1/2},F',M'_F\rangle)/\sqrt{2},
\end{eqnarray}
where on the right-hand side also $\Lambda'=0$ and $S'=1/2$ are implied.

By combining Eqs.\ (\ref{eq:tdm}), (\ref{eq:pi}), and (\ref{eq:sigma}), we obtain for the transition dipole moment
\begin{eqnarray}\label{eq:tdmfin}
\lefteqn{
\langle ^2\Sigma^+,F',M'_F,+| \mu^{\rm sf}_{{\rm el/mag},m}
|F_1 {^2\Pi}_{3/2},F,M_F,\epsilon\rangle =
\sqrt{(2 J +1)(2 J' +1)(2 F +1)(2 F' + 1)}
} \nonumber\\
&&\mbox{}\times
(-1)^{I-J+1-M_F'}
\begin{Bmatrix}
J & I & F \\
         F' & 1 & J' \\
  \end{Bmatrix}
  \begin{pmatrix}
        F & 1 & F' \\
        M_F & m & -M'_F \\
  \end{pmatrix}
\sum_{\Omega',\Omega,k}\frac{c_{\Omega}\epsilon^{\vartheta(\Omega)}}{2}
(-1)^{J'-\Omega'} {\mu}^{\rm bf}_{{\rm el/mag},k}
\begin{pmatrix}
        J & 1 & J' \\
        \Omega & k & -\Omega' \\
  \end{pmatrix},
\end{eqnarray}
where as before $J=3/2$, $I=1/2$, $J'=1/2$, and $I'=1/2$. Moreover, 
$\epsilon=\pm 1$, $\vartheta(x)$ is the step function, and we
need to sum over $\Omega'\in\{\pm 1/2\}$ and $\Omega\in\{\pm 1/2,\pm
3/2\}$. The factor $\epsilon^{\vartheta(\Omega)}$ contains
$\vartheta(-\Omega) = 0$ for $\Omega>0$ and $\vartheta(-\Omega) = 1$ for
$\Omega<0$. It is present because for $\epsilon=+1$ ($e$ %parity
symmetry) we have the positive combination in Eq.\ (\ref{eq:pief}), while for
$\epsilon=-1$ ($f$ %parity
symmetry) we have a sign change for negative $\Omega$.
The expression in Eq.\ (\ref{eq:tdmfin}) is valid for both the magnetic
and the electric transition. As follows from Eq.\ (\ref{eq:mu}), the two
transition dipole moments have a different $k$ dependence, since the
magnetic transition dipole moment changes sign when $k$ changes sign,
while the electric transition dipole moment does not. As a result, the
sum on the right-hand side of Eq.\ (\ref{eq:tdmfin}) leads to the
correct parity selection rules for the magnetic ($p = +\rightarrow+$)
and electric dipole transitions ($p = -\rightarrow+$).

\section{Comparison of theory with experiment}

In section \ref{sec:exp}, we distinguished between three different
experimental cases for the electric dipole transition, characterized by
the three different $M_F$ states of the $X\,^2\Pi_{3/2}, J=3/2, v= 0, e,
F=1$ state after the microwave transition. The electric dipole
transition strength to the $A\,^2\Sigma^{+}, J'=1/2, v'= 1$ state was
determined experimentally for these three cases separately. By the
symmetry of the experiment, case 2 with $M_F= 1$ and case 3
with $M_F= -1$ are equivalent. Moreover, since the experimental results
are in arbitrary units, we can thus compare only a single ratio of the
two inequivalent electric dipole intensities with theory. Since the
electric field of the laser is in the space-fixed $z$ direction, it
couples to the $z$ component of the space-fixed transition dipole
moment, so that $m=0$. Looking at Eq.\ (\ref{eq:tdmfin}), we find
that the two inequivalent cases only differ in the value of the
conserved $M_F$ and the final angular momentum $F'$, where only $F'=0$
gives a nonzero contribution for case 1 and only $F'=1$ for case 2 (and
3). We note that the part of Eq.\ (\ref{eq:tdmfin}) that depends on
$F'$, $M_F$, and $M_F'$ is given by
\begin{equation}
\xi({\rm case}) = \sqrt{2F'+1}(-1)^{-M'_F}\begin{Bmatrix}
         J & I & F \\
         F' & 1 & J' \\
  \end{Bmatrix}
  \begin{pmatrix}
        F & 1 & F' \\
        M_F & m & -M'_F \\
  \end{pmatrix},
\end{equation}
where $J=3/2$, $J'=1/2$, $I=1/2$, $F=1$ and $m=0$ for the
electric dipole transition. For case 1, we have $F'=0$ and $M_F=M_F'=0$,
while for case 2 we have $F'=1$ and $M_F=M_F'=1$. As a result, we find
the ratio between the two electric dipole transition strengths
analytically as the ratio of the squares of the two transition dipole
moments, resulting in
\begin{equation}\label{eq:anresult}
 {\rm signal \, ratio} =\frac{\xi({\rm case \, 1})^2}{\xi({\rm case \, 2})^2}=4.
\end{equation}
This is because all other terms of Eq.\ (\ref{eq:tdmfin}) drop out of
the ratio. Also other quantities as the laser intensity and the
molecular density drop out of the ratio because they are assumed to be
kept constant for the two experimental cases. The analytic result of
Eq.\ (\ref{eq:anresult}) was confirmed experimentally (see main text).

Next, we wish to calculate the ratio between the electric and the
magnetic dipole transition intensity. We note that the initial state for
the magnetic dipole transitions is the same for the three experimental
cases. Namely, it is given by a statistical mixture of $X\,^2\Pi_{3/2},
J=3/2, v= 0, f, F=2, M_F$ states, of which four levels $(M_F=\pm 1, \pm
2)$ are occupied by 20\% of the molecules and one level $(M_F=0)$ is
occupied by 19.5\%. For the electric transition, we define the following
line strength $S^{\rm el}$ for case 1, namely
\begin{eqnarray}\label{eq:sel}
S^{\rm el}&=& 0.005 | \langle ^2\Sigma^+,F'=0,M'_F=0,+|\mu^{\rm sf}_{{\rm el},0} |F_1 {^2\Pi}_{3/2},F=1,M_F=0,+\rangle|^2,
\end{eqnarray}
where the factor of 0.005 represents the 0.5 \% population of the
initial level and we have $\epsilon = +$ ($e$ symmetry or $-$ parity) for the
initial state. Since the magnetic field of the laser points in the $x$
direction of the laboratory frame, we define the following line strength
for the magnetic transition
\begin{eqnarray}\label{eq:smag}
S^{\rm mag}&=& \frac{\alpha^2}{2}\sum_{M_F} P(M_F)\sum_{M_F'} | \langle
^2\Sigma^+,F'=1,M'_F,+|(\mu^{\rm sf}_{{\rm mag},1}-\mu^{\rm sf}_{{\rm
mag},-1})|F_1 {^2\Pi}_{3/2}, F=2,M_F, - \rangle|^2,
\end{eqnarray}
where $P(M_F)$ accounts for the fraction of molecules in each initial
state, so that $P(M_F)=0.2$ for $M_F=\pm 1,\pm 2$ and $P(0)=0.195$,
and $\epsilon = -$ ($f$ symmetry or $+$ parity) for the initial state. Here,
$\alpha =1/137.036$ is the fine-structure constant, which is included to account for the relative strength of the magnetic
field of the laser compared to the electric field. Using Eqs.\
(\ref{eq:sel}) and (\ref{eq:smag}), we find for the ratio between the
electric and the magnetic dipole moment for case 1 that
\begin{eqnarray}
{\rm signal \, ratio}=\frac{S^{\rm el}}{S^{\rm mag}}=25.8.
\end{eqnarray}
This is to be compared with the experimental measured ratio of $18 \pm 8$.

Finally, we note that the above measured result depends on the fraction
of molecules in each of the prepared quantum states. If the two ground
states ($e$ and $f$) have an equal number of molecules before the dipole
transitions and if spatial orientation plays no role (for example in the
case of an unpolarized laser, or with all $M_F$ states equally
populated), then the signal ratio would have been simply given by
\begin{eqnarray}\label{eq:ratio}
\frac{1}{\alpha^2} \frac{|\langle \mu_{\rm el}\rangle|^2}{|\langle \mu_{\rm mag}|\rangle^2}=2576,
\end{eqnarray}
which therefore represents most directly the relative importance of
magnetic and electric dipole transitions from the $X\,^2\Pi_{3/2}, v= 0$
state to the $A\,^2\Sigma^{+}, v'= 1$ state.

\section{Discussion}

In the present theoretical treatment several approximations were made. First, we have separated the vibrational motion from the rotational motion. To partly compensate this approximation, we used a spectroscopic rotational constant, which effectively includes the effect of the vibrational motion, and we added a centrifugal term $J(J+1)/2\mu r^2$ to the electronic potentials when calculating the vibrational wave functions. Here, $r$ is the OH bond length and $\mu$ is the reduced mass of OH. If we excluded the centrifugal term, then the calculated signal ratios changed only by about 0.5\%, showing that these types of corrections are small. We also did not consider the $r$ dependence of the spin-orbit coupling, which affects the mixing of the $\Omega=1/2$ and the $\Omega=3/2$ states in the $^2\Pi$ state. The spin-orbit coupling varies less than 1\% in the range of interest to us. Moreover, we did not include a $\Lambda$-doublet Hamiltonian, which mixes in higher electronic states and causes the energy splitting between states of different parity in the $^2\Pi$ state. The resulting effect on the wave functions is very small, so that also the effect on the dipole transitions is expected to be small. Finally, we did not consider electric quadrupole transitions. An electric quadrupole transition can couple states with $\Delta F=\pm2$. Since in our experiment the laser propagates in the $y$-direction and is polarized in the $z$-direction, only the (space-fixed) $Q_{zy}$ component of the transition quadrupole moment is non-zero, see Eq.~(4.13.9) in \cite{craig:84}. This leads to the selection rule $\Delta M_F = \pm 1$ for our experimental situation. These transitions were not seen in the experiment. Electric quadrupole transition with $\Delta F=0,\pm1$ should be in the same order of magnitude as transition with $\Delta F = \pm2$ and can therefore be neglected. From calculations of the transition quadrupole moment with MOLPRO, we estimate that the quadrupole transitions are about $100$ times weaker than the magnetic dipole transitions, which is consistent with our experimental findings.  

Another theoretical issue is that the Franck-Condon overlap between the
$v=0$ and $v'=1$ is small and sensitive to shifts in the electronic
potentials. As an example of this, we note that the calculated ratio for
the $v=0 \rightarrow v'=1$ transition differs by almost a factor of 30
from the $v=1 \rightarrow v'=0$ transition. To get additional insight in
the sensitivity of our theoretical results to the method for calculating
electronic potentials, we re-calculated the $^2\Pi$ and $^2\Sigma$
potentials using the explicitly correlated RCCSD(T)-F12 method with an
aug-cc-pV5Z basis set \cite{molpro:10}.\\
In addition, we constructed RKR potentials using the program of Le Roy \cite{Roy} and the spectroscopic data of Crosley et al. \cite{Luque}. With the three independently calculated potentials we found a maximum deviation of only 3\% for the electric transition dipole moment and 2\% for the magnetic transition dipole moment. Moreover, we performed convergence checks of the electronic structure calculations for the dipole moment functions by changing the parameters in the configuration-interaction procedure that accounts for the effects of dynamic electron correlation. For the electric transition dipole moment this gave a maximum deviation of about 5\%. The magnetic transition dipole moment seemed to be better converged, with deviations of less than 1\%. Combining these errors we estimate an uncertainty of about 10\% for the ratio of the squared transition dipole moments. This estimate is only crude, however. The convergence of configuration-interaction calculations is typically not very regular. Furthermore, systematic errors in the calculation of the potentials, such as the Born-Oppenheimer approximation, could influence the Franck-Condon factor.
%Using these independently calculated potentials we did not find significant deviations from the results obtained with the potentials by Van der Loo (only a few \% difference). However, it could still be that both methods suffer from similar sytematic errors that possibly influence the Franck-Condon factor, such as the Born-Oppenheimer approximation for example.

% \bibliography{../vanderwaals}